\documentclass[letterpaper,twocolumn,english,aps,prl,superscriptaddress]{revtex4}
\usepackage{lmodern}
\usepackage{lmodern}
\usepackage[LGR,T1]{fontenc}
\usepackage[latin1]{inputenc}
\usepackage{color}
\usepackage{textcomp}
\usepackage{amsbsy}
\usepackage{amstext}
\usepackage{amssymb}
\usepackage{wasysym}
\usepackage{graphicx}
\usepackage{esint}

\makeatletter

\@ifundefined{textcolor}{}
{%
 \definecolor{BLACK}{gray}{0}
 \definecolor{WHITE}{gray}{1}
 \definecolor{RED}{rgb}{1,0,0}
 \definecolor{GREEN}{rgb}{0,1,0}
 \definecolor{BLUE}{rgb}{0,0,1}
 \definecolor{CYAN}{cmyk}{1,0,0,0}
 \definecolor{MAGENTA}{cmyk}{0,1,0,0}
 \definecolor{YELLOW}{cmyk}{0,0,1,0}
}

\usepackage{amsbsy}
\usepackage{amstext}

\usepackage{epsfig}
\usepackage{dcolumn}
\usepackage{bm}
\usepackage{bbm}

\usepackage{blindtext}

\usepackage{tikz}

\usepackage{babel}

\makeatother

\usepackage{babel}

\begin{document}

\title{\textcolor{black}{Designing Quantum Spin-Orbital Liquids in Artificial Mott Insulators}}

\author{Xu Dou}

\affiliation{Department of Physics and Astronomy, University of Oklahoma, Norman,
OK 73069, USA}

\author{Valeri N. Kotov}

\affiliation{Department of Physics, University of Vermont, Burlington, VT 05405,
USA }

\author{Bruno Uchoa}


\selectlanguage{english}%

\affiliation{Department of Physics and Astronomy, University of Oklahoma, Norman,
OK 73069, USA }

\date{\today}

\begin{abstract}
Quantum spin-orbital liquids are elusive strongly correlated states of matter that emerge from
quantum frustration between spin and orbital degrees of freedom. 
A promising
route towards the observation of those  states is the creation of
artificial Mott insulators where antiferromagnetic correlations between
spins and orbitals can be designed. 
We show that Coulomb impurity lattices on the surface of gapped honeycomb
substrates, such as graphene on SiC, can be used to simulate SU(4) symmetric spin-orbital lattice models. 
We exploit the property that massive Dirac fermions 
form mid-gap bound states with spin and valley degeneracies in the vicinity of a Coulomb impurity.
Due to electronic repulsion, the antiferromagnetic correlations of the impurity lattice are driven by a super-exchange interaction with SU(4) symmetry, which emerges from the  bound states degeneracy at quarter filling. We propose that quantum spin-orbital liquids can be engineered in artificially designed solid-state systems at vastly higher temperatures than achievable in optical lattices with cold atoms. We discuss the experimental setup and  possible scenarios for candidate quantum spin-liquids in Coulomb impurity lattices of various  geometries. 

\end{abstract}


\maketitle


Quantum spin liquids are highly entangled states that can emerge in
antiferromagnetic lattices in the presence spin degeneracies and frustration.
Spin-orbital liquids result from systems that have not only spin degeneracies
but also orbital degeneracies \cite{Lee,Balents}. 
Those states 
are strongly correlated, have non-local excitations, but nevertheless do not break any symmetries. 
In spite of mounting
theoretical effort \cite{Kitaoka,Mostovoy,Reitsma,Fritsch,Vernac},
a significant difficulty in finding viable candidates for quantum
spin-orbital liquids is the fact that normally the interactions governing
spin and orbital degrees of freedom have very different energy scales
\cite{Bonda,Fichtl,Lewis}. Consequently those degrees of freedom
are decoupled at sufficiently low temperatures, hindering the quantum
frustration that is required to entangle orbitals and spins. Very
recently, x-ray scattering studies in magnetic honeycomb based BaCuSb$_{2}$O$_{9}$
crystals reported indications of spin-orbital entanglement at low
temperature \cite{SO liquid1,Ishiguro}. 

An alternative to identifying crystals where spins and orbtitals are
strongly coupled would be instead to create artificial crystals where
spin and orbital quantum numbers become interchangeable. Such property
appears in magnetic Hamiltonians that display SU(4) symmetry \cite{Li}.
Recent experiments with cold atoms reported spectroscopic quantum
simulations in small artificial magnetic systems with SU($N\leq10$) symmetry 
 at ultra low temperature 
 \cite{Gorshov, zhang}. Mott physics with SU(2) spins has been observed in 
 optical lattices with ultra cold atoms inside a parabolic potential \cite{Jordens}. 
In those systems, strong correlations emerge only at extremely low temperatures, 
making a possible detection of quantum spin-liquids challenging \cite{Cazalilla}.
 Solid-state systems where
antiferromagnetic interactions have SU(4) symmetry are not common,
since in real materials, anisotropies and off-diagonal hopping matrix
elements in the degenerate orbital space usually lower that symmetry
\cite{kugel}. 

We propose a solid-state system that can be experimentally designed
with scanning tunneling microscopy (STM) tips by positioning Coulomb
impurity adatoms in a periodic array on top of an insulating honeycomb
substrate. The electrons in those substrates can be described by massive
Dirac fermions, which form bound states around the impurities \cite{Pereira,Novikov,Kotov}.
Those bound states have spin and valley degeneracies, which are dual
to spin-orbital degrees of freedom. We theoretically construct an
artificial lattice where each impurity site is quarter filled with
valley and spin polarized states. The problem has an emergent SU(4)
symmetry that follows from the orthogonality between the two different
valley spaces. In systems like graphene, SU(4) symmetry is known to 
emerge in the quantum Hall regime \cite{Goerbig}. Electronic interactions lead to a variety 
of broken symmetry states in both spins and valleys \cite{Young,Young2, Nomura, Alicea, Abanin, Sodermann}.

The spin-orbital exchange interactions are calculated in three different
impurity lattice geometries: triangular, square and honeycomb, shown
in Fig. 1. We find the constraints on the impurity lattice in the
regimes where the system is expected to behave as a Mott insulator
dominated by antiferromagnetic interactions between sites. We propose
the experimental conditions for the observation of those states. For
honeycomb substrates such as graphene grown on SiC \cite{Lanzara,Nevius},
we show that the Mott regime of entangled spins and orbitals is
experimentally accessible and that the superexchange interaction can
be as large as $J_{s}/k\sim60-120$ K. The experimental signatures
of strongly correlated states are discussed based on possible scenarios
predicted for SU(4) spin-orbital models \cite{Corboz,Penc,Corboz-1,Wang},
including quantum spin-orbital liquids. 

\begin{figure*}
\includegraphics{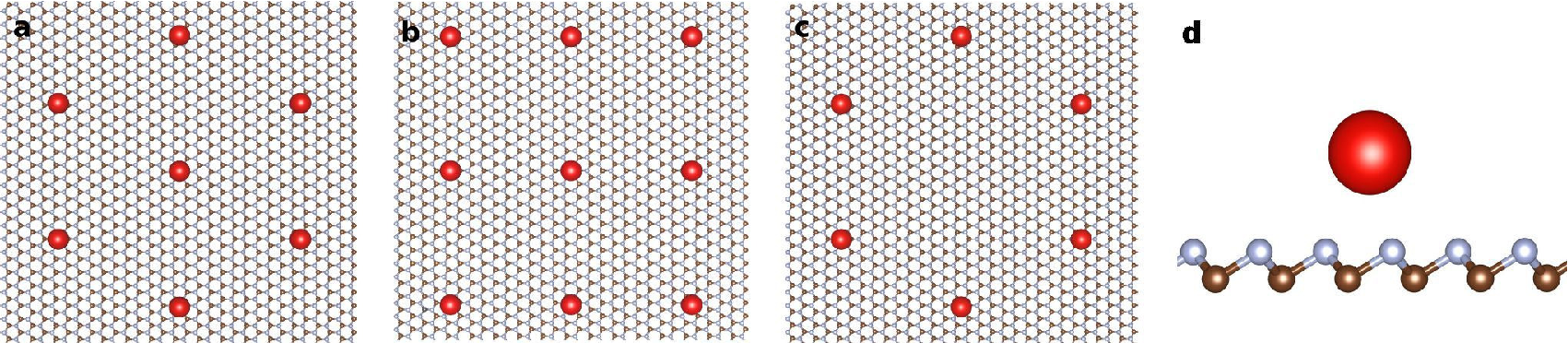}

\protect\caption{ {\bf Coulomb impurity lattices. }Honeycomb substrate with unequal
sublattices decorated with a superlattice of charged impurities. In
the three configurations, triangular ({a}), square ({b}) and honeycomb ({c}), the impurities
are separated by a superlattice\textbf{ }constant $L$, and sit at
a distance $d$ away from the plane of the substrate ({d}). All impurities
interact with electrons via Coulomb, $1/r$ potential. }
\end{figure*}

\section*{Results}

\textbf{Coulomb impurity problem.} The wavefunction of the Coulomb
impurity bound states for 2D massive Dirac fermions, $\Psi(\mathbf{r})$,
can be derived from the Dirac equation 
\begin{equation}
(-i\hbar v\boldsymbol{\sigma}\cdot\boldsymbol{\nabla}+V(r)+mv^{2}\sigma_{z})\Psi(\mathbf{r})=\epsilon\Psi(\mathbf{r}).\label{eq:Dirac}
\end{equation}
$\boldsymbol{\sigma}=(\sigma_{x},\sigma_{y})$ is a vector with off-diagonal
Pauli matrices, $\sigma_{z}$ is the diagonal Pauli matrix, $v$ is
the Fermi velocity and $m$ is the mass term of the substrate, that
describes a gap in the electronic spectrum, $\Delta=2mv^{2}$. $V(r)=-Ze^{2}/\kappa\sqrt{r^{2}+d^{2}}$
is the Coulomb impurity potential, where is $Z$ the atomic number
of the impurity, $e$ is the electron charge, $\kappa$ the dielectric
constant of the surface, and $d\approx2-3\mbox{\AA}$ is the out-of-plane
distance between the impurity and the plane of the substrate.

The impurity potential decays as $V(r)\sim1/r$ in the $r\gg d$ limit
and saturates into a constant in the opposite limit. The potential
can be written as 
\begin{equation}
V(r)=-Z\frac{e^{2}}{\kappa}\left[\frac{1}{r}\theta(r-a)+\frac{1}{a}\theta(a-r)\right]\label{eq:Vr}
\end{equation}
where $a$ is an effective real space cut-off which regularizes the
Coulomb potential. The size of the cut-off can be chosen as $a\sim d$
and is typically of the order of the impurity size. This regularization
procedure is well known in quantum electrodynamics in 3+1 dimensions
(QED$_{3+1}$) and has been successfully used to explain the experimentally
observed dive of bound states in the lower continuum around super-heavy
nuclei with atomic number $Z>137$ \cite{Greiner,Zeldovich}. Both
in QED$_{3+1}$ as in the 2D case, the wavefunction of the Coulomb
impurity bound states decay over a characteristic distance defined
by the Compton wavelength $\lambda_{C}=\hbar/mv$. 

In cylindrical coordinates, the solution of Eq. (\ref{eq:Dirac})
is in the form 
\begin{equation}
\Psi(r,\phi)=\frac{c}{\sqrt{2\pi}}\left(\begin{array}{c}
F_{j}^{(-)}(r)\mbox{e}^{i(j-\frac{1}{2})\phi}\\
iF_{j}^{(+)}(r)e^{i(j+\frac{1}{2})\phi}
\end{array}\right),\label{eq:phi}
\end{equation}
where $j=\pm\frac{1}{2},\,\pm\frac{3}{2}\,\ldots,\, m+\frac{1}{2}$
($m\in\mathbb{Z}$) are the possible angular momentum states, and
$c$ is the normalization constant. The energy spectrum is quantized
by the usual quantum numbers in the Hydrogen atom problem, $n\in\mathbb{N}$
and $j$ \cite{Pereira,Novikov,Kotov}. The degeneracy of the $\pm|j|$
angular momenta states for a given $n>0$ however is lifted. At $n=0$,
only the $j=\frac{1}{2}$ state is allowed. 

Defining the impurity strength by the dimensionless coupling $g\equiv Z\alpha$,
where $\alpha=e^{2}/\kappa\hbar v$ is the screened fine structure
constant of the substrate, there are two known regimes of the problem:
the perturbative regime $g\ll1$, where the bound states are shallow,
and the strong coupling regime $g\gtrsim0.5$, where they dive in
the negative sector of the energy spectrum, as shown in Fig. 2. At
fixed $g$, the lowest energy level is the $n=0$, $j=\frac{1}{2}$
state, followed by the first excited state $n=1$, $j=-\frac{1}{2}$.
There is an infinite number of higher excited states inside the gap
$\Delta$. \textcolor{black}{The latter states have very small binding
energies and are not relevant to this discussion. }

We are interested in the strong coupling regime of the problem ($g\gtrsim0.5$),
where the confining potential is deep and the energy separation between
the ground state level and the first excited state is of the order
of $\sim\Delta/2$. At sufficiently large coupling, $g>g_{c}$, the
lowest energy state level dives in the continuum of negative energy
states outside of the gap. This regime is known as the supercritical
regime. At the critical one, when $g=g_{c}$ the energy of the lowest
level is exactly at the edge of the gap, $\epsilon=-mv^{2}$. In the
subcritical regime, $0.5\lesssim g<g_{c}$, which is the focus of
this paper, the levels are strongly localized and sharply defined
inside the gap. For a Coulomb impurity on top of graphene epitaxially
grown on SiC, where $\Delta\sim0.26$ eV \cite{Lanzara}, and for
a typical small distance cut-off $a\approx2.8\mbox{\AA}$, $g_{c}=0.916$. In
general, the critical coupling $g_c \sim 1$.  
The energy of the levels follows directly from matching the wave function
at $r=a$, similarly to the procedure in the QED$_{3+1}$ case. The
solution of the subcritical regime can be calculated either numerically
\cite{Pereira} or for the purposes of this work, analytically, as detailed in the Supplemental Materials. 

\textbf{Impurity lattice model.} In a honeycomb lattice with massive Dirac
fermions, the quasiparticles also have two valley flavors, in addition
to the spin. The Coulomb impurity bound states therefore must have
both spin and valley degrees of freedom. The Dirac equation in this
case is 
\begin{equation}
\left(\begin{array}{cc}
\hat{\mathcal{H}}_{+}(\mathbf{r}) & 0\\
0 & \hat{\mathcal{H}}_{-}(\mathbf{r})
\end{array}\right)\Phi(\mathbf{r})=\epsilon\Phi(\mathbf{r}),\label{eq:PhiDirac Eq}
\end{equation}
where $\hat{\mathcal{H}}_{+}(\mathbf{r})=-i\hbar v\boldsymbol{\sigma}\cdot\boldsymbol{\nabla}+V(r)+mv^{2}\sigma_{z}$
is the Dirac Hamiltonian matrix in valley $+$ and $\hat{\mathcal{H}}_{-}(\mathbf{r})=\hat{\mathcal{H}}_{+}^{*}(\mathbf{r})$
in the opposite valley. The eigenvectors are the four component spinors
$\Phi_{j,+}(\mathbf{r})=(\Psi_{j}(\mathbf{r}),\mathbf{0})$ and $\Phi_{j,-}(\mathbf{r})=(\mathbf{0},\Psi_{j}^{*}(\mathbf{r}))$,
which are degenerate. The $j$-th energy level is four-fold degenerate,
with two spins and two valleys. The valleys describe the orbital motion
of an electron around a Coulomb impurity. They effectively behave
as a pseudo-spin with SU(2) symmetry, as the actual spins.

\begin{figure}[t]
\begin{centering}
\includegraphics[scale=0.36]{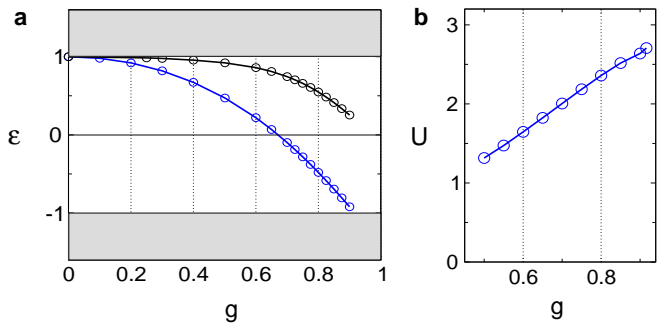} 
\par\end{centering}

\protect\protect\caption{ {\bf Single impurity energy scales. } ({a}) Energy of the Coulomb
impurity bound states $\epsilon$, in units of $mv^{2}=0.13$ eV,
as a function of the dimensionless coupling $g=Z\alpha$. Blue dots:
ground state level, $n=0$, $j=\frac{1}{2}$. Black dots: first excited
state, $n=1$, $j=-\frac{1}{2}$. At $g=g_{c}\approx0.916$, the lowest
energy level dives in the continuum of negative energy states at $\epsilon=-mv^{2}$.
In the subcritical regime $g\lesssim g_{c}$, the two levels have
an energy separation $\sim mv^{2}$. ({b}) Hubbard $U$, in units
of $mv^{2}\alpha$, versus $g$ in the strong coupling regime $0.5\leq g\leq g_{c}$.
$U$ is comparable to the energy of the gap $\Delta=2mv^{2}$.}
\end{figure}

Once Coulomb interactions among the electrons in the bound state are
included, those states tend to spin and valley polarize due to correlations
and Pauli blocking. In the ground state, $j=\frac{1}{2}$, the Coulomb
interaction can be expressed in terms of a Hubbard $U$ term 
\begin{equation}\mathcal{H}_{U}=\frac{1}{2}U\sum_{\{\nu\},\{\sigma\}}\hat{n}_{\nu,\sigma}\hat{n}_{\nu^{\prime},\sigma^{\prime}}(1-\delta_{\nu,\nu^{\prime}}\delta_{\sigma\sigma^{\prime}}),
\end{equation}
where 
\begin{equation}
U=\int\mbox{d}^{2}r\mbox{d}^{2}r^{\prime}|\Phi_{\frac{1}{2},\nu}(\mathbf{r})|^{2}\frac{e^{2}}{\kappa|\mathbf{r}-\mathbf{r}^{\prime}|}|\Phi_{\frac{1}{2},\nu^{\prime}}(\mathbf{r}^{\prime})|^{2}\label{eq:U}
\end{equation}
is a valley independent local repulsion. $\hat{n}_{\nu,\sigma}=c_{\nu,\sigma}^{\dagger}c_{\nu\sigma}$
is the number operator per valley and spin at the bound state, where
$c_{\nu,\sigma}$ annihilates one electron in the $j=\frac{1}{2}$
level on valley $\nu$ with spin $\sigma$. Due to the orthogonality
of the eigenspinors, $\Phi_{j,+}^{\dagger}(\mathbf{r})\Phi_{j,-}(\mathbf{r})=0$,
the exchange interaction between electrons in different valleys around
the same Coulomb impurity is zero.

\begin{figure*}
\includegraphics[scale=0.5]{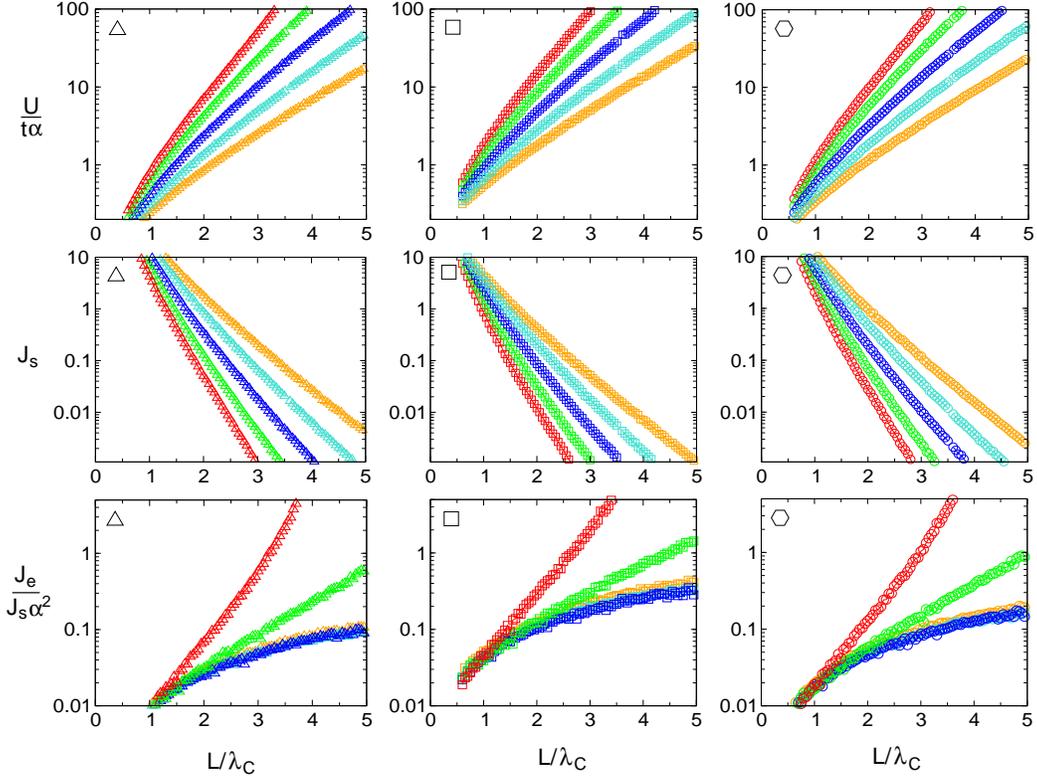}

\protect\protect\caption{ {\bf Correlations in Coulomb impurity lattices. }Left column $(\vartriangle)$:
triangular lattice; middle column $(\Square)$ square lattice; right
column $(\hexagon)$: honeycomb lattice. Red dots: $g=0.9$; green:
$g=0.8$; blue: $g=0.7$; cyan: $g=0.6$; orange: $g=0.5$. Top row:
ratio between the onsite repulsion $(U)$ and the kinetic energy $(t)$
times the fine structure $\alpha$ versus the superlattice constant
$L$ normalized by the Compton wavelength $\lambda_{C}=\hbar/mv$.
For a substrate with a gap of of \textcolor{black}{$\Delta=0.26$
eV (graphene on SiC),} $\lambda_{C}\approx46\mbox{\AA}$. When $U/t\gtrsim5$,
the system is strongly correlated: the Coulomb impurities form a lattice
of local spin-orbitals. Middle row: Superexchange interaction, $J_{s}=t^{2}/U$
in units of $mv^{2}/\alpha$, versus $L/\lambda_{C}$. Bottom row:
ratio between the exchange interaction $J_{e}$ and the superexchange
interaction $J_{s}$ times $\alpha^{2}$. }
\end{figure*}

In Fig. 2, we calculate $U$ as a function of the dimensionless impurity
coupling $g$ in the strong coupling regime $0.5\lesssim g<g_{c}$.
At $g=g_{c}$, $U=2.7\, mv^{2}\alpha$, dropping to $U=1.35\, mv^{2}\alpha$
at $g=0.5$. When $U$ is large and only the $j=\frac{1}{2}$ level
is filled, the ground state will be singly occupied in one of the
four possible states: $|\,$\tikz\draw[red,fill=red] (0,0) circle (.6ex);$\,\rangle\:=|+,\uparrow\rangle$,
$|\,$\tikz\draw[blue,fill=blue] (0,0) circle (.6ex);$\,\rangle\:=|+,\downarrow\rangle$,
$|\,$\tikz\draw[green,fill=green] (0,0) circle (.6ex);$\,\rangle\:=|-,\uparrow\rangle$,
and $|\,$\tikz\draw[yellow,fill=yellow] (0,0) circle (.6ex);$\,\rangle\:=|-,\downarrow\rangle$.

We would like to write down an effective lattice model for a strongly
correlated lattice of Coulomb impurities, each one having a quarter
filled bound state in one of the four possible states above. Those
electrons can hop between different Coulomb impurity sites, with each
one having a Hubbard $U$ energy, that penalizes multiply occupied
sites, and also having a well defined valley and spin. The hopping term is
\begin{equation}
\mathcal{H}_{t}=-t\sum_{\langle ij\rangle}\sum_{\nu,\sigma}c_{i,\nu,\sigma}^{\dagger}c_{j,\nu,\sigma},
\end{equation}
with $c_{i}$ describing the annihilation operator of an electron
in the $j=\frac{1}{2}$ level siting on an impurity site located at
$\mathbf{R}_{i}$, and $\langle ij\rangle$ denotes summation over
nearest neighbor (NN) sites.  
The hopping parameter of the Coulomb impurity lattice is 
\begin{equation}
t_{ij}=\int\mbox{d}^{2}r\Phi_{\frac{1}{2},\nu}^{\dagger}(\mathbf{r}_{i})\sum_{k\neq i}V(|\mathbf{r}_{k}|)\Phi_{\frac{1}{2},\nu}(\mathbf{r}_{j})\label{eq:t}
\end{equation}
where $\mathbf{r}_{i}\equiv\mathbf{r}-\mathbf{R}_{i}$ is the position
relative to site $i$.
Hopping between Coulomb impurity sites conserves valley due 
to the orthogonality of eigenspinors in the valley space, $\Phi^\dagger_{\frac{1}{2},+}(\mathbf{r}_i)\Phi_{\frac{1}{2},-}(\mathbf{r}_j)=0$.
Because of the summation of the potential over
lattice sites and the long range nature of the Coulomb interaction,
the value of $t$ is influenced by the geometry of the lattice.

In the limit $U\gg t$, we can expand the effective Hamiltonian in
second order perturbation theory in the hopping, $\mathcal{H}_{s}=-\mathcal{H}_{t}\mathcal{H}_{U}^{-1}\mathcal{H}_{t}+\mathcal{O}(t^{4})$.
The Hamiltonian that results is the superexchange interaction, which
favors antiferromagnetic alignment of spins or valleys. This interaction
is of order $J_{s}=t^{2}/U$ and lowers the energy cost for electrons
to hop back and forth between two NN sites. The superexchange competes
with the exchange interaction between NN sites, which is ferromagnetic
and defined by $J_{e,ij}=-\frac{1}{2}\int\mbox{d}^{2}r\mbox{d}^{2}r^{\prime}\,\Phi_{\frac{1}{2},\nu}^{\dagger}(\mathbf{r}_{i})\Phi_{\frac{1}{2},\nu}(\mathbf{r}_{j})\frac{e^{2}}{\kappa|\mathbf{r}-\mathbf{r}^{\prime}|}\Phi_{\frac{1}{2},\nu^{\prime}}^{\dagger}(\mathbf{r}_{j}^{\prime})\Phi_{\frac{1}{2},\nu^{\prime}}(\mathbf{r}_{i}^{\prime}),$
with $J_{e,\langle ij\rangle}\equiv J_{e}<0$. As shown in the Methods section, 
both the superexchange and the exchange interactions map
into a Kugel-Khomskii type Hamiltonian \cite{Kugel} with \emph{exact}
SU(4) symmetry, 
\begin{equation}
\mathcal{H}=J\sum_{\langle ij\rangle}\left(\frac{1}{2}+2\boldsymbol{\tau}_{i}\cdot\boldsymbol{\tau}_{j}\right)\left(\frac{1}{2}+2\mathbf{S}_{i}\cdot\mathbf{S}_{j}\right),\label{eq:JHam}
\end{equation}
where $\boldsymbol{\tau}_{i}$ is the valley pseudospin operator and
$\mathbf{S}_{i}$ the spin operator on a given site. Hamiltonian
(\ref{eq:JHam}) is symmetric under any permutation among the four
different valley-spin flavors (colors).

The coupling $J\sim J_{s}>0$ in the regime where the superexchange
coupling dominates ($t^{2}/U\gg J_{e}$). The superexchange interaction
is antiferromagnetic, and can drive the spin-orbital lattice into
frustrated phases where no symmetry is broken. In the opposite regime
($J_{e}\gg t^{2}/U$), the coupling $J=-J_{e}<0$ changes sign, and
the system tends to order in a ferromagnetic state at zero temperature.

\textbf{Numerical results. }In Fig. 3 we show the ratio of $U/t\alpha$
as a function of the impurity lattice constant $L$ for three different
geometries: triangular ($\vartriangle$), square ($\Square$) and
honeycomb ($\hexagon$). $L$ is normalized by the Compton wavelength
$\lambda_{C}$, which is inversely proportional to the mass gap of
the substrate. In the regime where $U/t\gtrsim5$, the system is a
strongly correlated insulator and can be effectively described as
a lattice of local valley-orbitals and spins. The different curves
in each panel correspond to different impurity couplings, with $g$
ranging from $0.5$ to the critical value $g_{c}\sim0.916$. At the
middle row panels, we display the superexchange coupling $J_{s}$
(in units of $mv^{2}/\alpha$) as a function of $L$. For couplings
$g<g_{c}$, when $U/t\alpha\sim12$ the superexchange coupling ranges
from $J_{s}\alpha/mv^{2}\approx0.01-0.02$ for $g$ running between
0.5 and 0.9 in all geometries we tested, as indicated in Fig. 3. In the regime
$U/t\alpha\sim20$, the super exchange is in the range $J_{s}\alpha/mv^{2}\approx 0.003 - 0.007$.

For graphene on SiC substrate with $\Delta=2mv^{2}\sim0.26$ eV, the
Compton wavelength $\lambda_{C}\approx46\mbox{\AA}.$ On the surface
of SiC ($\kappa\sim5.2$) the fine structure constant $\alpha\approx0.42$.
The size of the superlattice constant $L$ that corresponds to a fixed
value of $J_{s}$ varies slightly depending on the geometry of the
lattice. At $g\approx g_{c}$ (red dots), the impurity valence $Z\sim2$.
When $U/t\alpha=12$ ($U/t\approx5$), the superexchange interaction
between NN sites is $J_{s}/k\sim59$ K and corresponds to impurity
lattice constants $L/\lambda_{C}\approx2.25\,(\vartriangle),\,1.9\,(\Square),$
and $2.1\,(\hexagon)$, resulting in $L\sim90-100\mbox{\AA}$. At
$g=0.5$ or $Z\sim1$ (orange dots), the wavefunctions are more weakly
bounded to the impurities and hence more extended. The same ratio
of $U/t\approx5$ corresponds to $J_{s}/k\sim28$ K and larger superlattice
constants $L/\lambda_{C}\sim4.6\,(\vartriangle),\,3.9\,(\Square),\,$and
$4.3\,(\hexagon)$, respectively, with $L\sim180\mbox{\AA}-200\mbox{\AA}$.
For a larger gap of $\Delta\sim0.5$ eV \cite{Nevius}, the superexchange
nearly doubles ($J_{s}\sim56-118$ K) while the Compton wavelength
is halved. When $U/t\alpha = 20$ ($U/t\approx 8.5$),  $J_s/k \sim 10-20$ K. 

In the regime of interest, where $U/t$ is large, $U$ is the largest
energy scale in the problem. The superexchange interaction competes
with the exchange one $J_{e}$ and, in principle, both can be of the
same order. In the bottom row of the panels in Fig. 3 we plot the
ratio between $J_{e}/J_{s}\alpha^{2}$. For $\alpha<1$, the superexchange
interaction clearly dominates the exchange interaction, and is at
least three times larger for $U/t\alpha\lesssim20$. When considering
Coulomb impurities on graphene-SiC substrates, where $\alpha=0.42$,
the ratio $J_{e}/J_{s}<0.07$ in all geometries considered in the
range $U/t\lesssim8.5$. The dominant interactions are therefore clearly
antiferromagnetic. Due to the SU(4) symmetry, valley and spin degrees
of freedom are strongly entangled and may give rise to a spin-orbital
liquid in the Mott insulator regime. 

\textbf{Experimental setup.} The lattice of Coulomb impurities can
be experimentally created with STM tips, which can drag atoms on a
surface with atomic precision \cite{Eigler}. Possible substrates
include graphene epitaxially grown on SiC, which was shown to develop
a gap ranging from $\Delta=0.26-0.5$ eV \cite{Lanzara,Nevius}. In high quality samples, 
the Fermi level was observed in the middle of the gap \cite{Nevius}. Other
crystals, such as MoS$_{2}$, MoSe$_{2}$, and other dichalcogenides
\cite{Heinz}, have even larger gaps, however they also exhibit large
spin-orbit couplings \cite{Zhu}, which will lift the SU(4) symmetry
of the problem, lowering it to SU(2). Strong unitary disorder  connects 
the  two valleys  and  can also have a similar effect. Disorder effects, however, 
can be inhibited by properly annealing the substrate.   

Among alkaline metals, potassium adatoms ($Z=1$) are known to spontaneously
form two dimensional crystals on honeycomb substrates such as graphite
\cite{Caragiu}. Higher valence cobalt adatoms have already been studied
with STM on graphene and are also possible candidates \cite{Brar}.\textcolor{blue}{{}
}\textcolor{black}{The strong coupling regime, where the bound states
are deep and well separated, is experimentally accessible for impurities
with a valence $Z\sim1$. That contrasts with the standard relativistic
scenario, where the strong coupling regime can be achieved only when
the valence is of the order of the inverse of the QED fine structure
constant $Z\sim1/\alpha_{QED}=137$. }

The determination of the impurity lattice constant $L$ that is required
to create a Mott insulator with strong antiferromagnetic correlations
can be achieved with local spectroscopy measurements around a single
impurity. Those measurements can accurately determine the energy of
the bound states inside the gap. With the theoretical wavefunctions,
one can extract the effective impurity coupling $g$ by comparing
the measurement of the energy levels with the calculated result, as
shown in Fig. 2. The appropriate range for the impurity lattice constant
is indicated in the plots of Fig. 3. Integration of the measured local
density of states over the area around the impurity gives the occupation
of the ground energy level inside the gap. When the impurity lattice
is in the Mott regime, each four-fold degenerate impurity level will
remain singly occupied (quarter filling). 

\section*{Discussion}

Recent numerical evidence \cite{Corboz}
suggests that the ground state of the antiferromagnetic Hamiltonian
(\ref{eq:JHam}) in the honeycomb lattice is a strongly correlated
state that preserves all the symmetries of the system. This state
is a quantum spin-orbital liquid schematically drawn in the left panel
of Fig. 4. Every site has a well defined spin-valley state (color)
among the four possible colors. Each color has the same neighbors
up to color permutations. The pattern preserves both the lattice symmetry
and the SU(4) color symmetry. 

Color-color correlations appear to decay
as a power law, indicating a gapless state, or equivalently, an algebraic
quantum spin-orbital liquid with no true long range order. \textcolor{blue}{{} }\textcolor{black}{Algebraic
spin liquids are generally known to be robust two-dimensional interacting
critical states, relevant to a variety of correlated physical models
\cite{Hermele}}\textcolor{blue}{.} 
After comparison of the energy
of several different states, the quarter filled $\pi$-flux state
currently appears as the leading candidate \cite{Corboz}. 
In the honeycomb lattice, a $\pi$-flux in the honeycomb
plaquette creates Dirac fermions at quarter filling, which is the
regime of interest for Mott insulators with SU(4) symmetry. Those
Dirac fermions are \textcolor{black}{(color)} spinon excitations,
which are four-fold degenerate due to the color symmetry.

Low-energy characteristic probes amenable to 2D systems
have been proposed, such as injecting a spin current into the insulator
and monitoring the spin bias dependence of the current \cite{Chatterjee,Chen}.
In the simplest experimental setup with a single
metal-insulator interface, spin accumulation is achieved via the spin
Hall effect. In the four-terminal
setup, the spin-liquid insulator is coupled to left and right metal
leads. Spin current detection occurs through the reverse spin Hall
effect in one of the metallic contacts. 

In the spin-orbital (valley) case at hand,
the spin degrees of freedom in
the insulator and in the metal are coupled at the interface. The valleys
are decoupled from the orbital degrees of freedom in the metal. Hence
the valleys do not experience flips due to the spin current injection.
The result is the propagation of a pure spin current with additional
valley degeneracy. Consequently, in this case, the spin current will
scale in the same way with the bias voltage as in pure spin models. 
For the $\pi$-flux state, the Dirac cone of the spinons is degenerate
in all quantum numbers (spin and valley). The spin current scales
with the fifth power of the bias voltage, $I_{s}\sim V^{5}$\textcolor{black}{{}
\cite{Chatterjee,Chen}.} This result appears to be a universal signature
of both spin and spin-orbital liquid phases with gapless Dirac fermion
spinons. In general, the power of the spin
voltage dependence is sensitive to the nature and dispersion of the
spinon excitations. The exact nature of the spin-orbital liquid state
in the honeycomb lattice requires further investigation. Nevertheless,
the prospects of observing a true quantum spin-orbital liquid in this
geometry seem quite promising.

\begin{figure}
\includegraphics[scale=0.36]{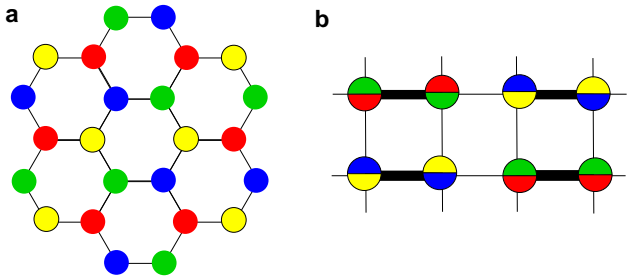}

\protect\protect\caption{ {\bf Spin-orbital color states. }({a}) Possible algebraic quantum
spin-orbital liquid for the honeycomb lattice in the SU(4) Heisenberg
model, numerically predicted in Ref. {[}\onlinecite{Corboz}{]}.
This state may correspond to a quarter filled $\pi$-flux phase. Each
color is surrounded by the same states, up to color permutations.
Both crystalline and SU(4) symmetries are intact. ({b}) Possible
dimerized state in the square lattice, with alternating singlets of
two colors (after Ref. {[}\onlinecite{Corboz-1}{]}.) This state
has long range order and breaks both lattice and color symmetry.}
\end{figure}

Triangular lattices are natural candidates for quantum
disordered states due to their strongly frustrated nature. It was
proposed at first that their ground state has plaquette order \cite{Li},
with plaquettes formed by SU(4) singlets. However more recent work
\cite{Penc} found strong local resonances between plaquette configurations.
While more complicated orders with large unit cells can not be ruled
out, the ground state appears to be a spin-orbital liquid with no
plaquette order. 
The presence of next-nearest neighbor superexchange
$J_{s}^{\prime}$ drives the system into magnetically long range ordered
state via a quantum phase transition at a critical value $J_{s}^{\prime}/J_{s}\approx0.12$
\cite{Penc}. In the proposed Coulomb impurity lattice, we find that
ratio to be $\sim10^{-2}$. On the basis of the existing knowledge
about the model, we conclude that a spin-orbital liquid state can
be realized in the Mott regime. The nature of this state is
not yet known.

There have been suggestions of a variety
of different ground states for Hamiltonian (\ref{eq:JHam}) in the
square lattice. Possibilities include a gapless spin liquid with nodal
fermions \cite{Wang}, and a plaquette state \cite{Li,Hung}. A more
recent numerical work has laid more concrete evidence towards a dimerized
state depicted in Fig. 4, which breaks both lattice and color symmetry
\cite{Corboz-1}. The thick bonds represent strong bonds, while the
think lines are weaker.
This particular state 
has two sets
of dimers with two colors each, which alternate along the two main
directions of the lattice. Because of the broken
symmetry, the elementary excitations are Goldstone modes in the form
of gapless (color) magnons. These could also lead to characteristic
power law dependencies in the spin current as a function of spin bias
\cite{Chatterjee,Chen}, with the power being generally smaller than
for gapless Dirac spinons ($\pi$-flux phase). 

 Coulomb impurity lattices offer wide possibilities
for different frustrated scenarios due to the inherent flexibility
in their design. Recent experiments observed evidence for a spin-liquid
ground state in the antiferromagnetic Kagome lattice \cite{Fu}. We
conjecture that gapped honeycomb substrates with large spin orbit
coupling, such as MoS$_{2}$ \cite{Zhu},  could
be experimentally used to design frustrated artificial Coulomb impurity
lattices where the spin degeneracy is explicitly lifted, leaving a
pure quantum orbital (valley) liquid in the ground state. 
The tendency towards frustration is not the unique
scenario for artificial lattices supported on gapped honeycomb substrates.
For instance, color ferromagnetism is possible in superlattices
of mass defects forming quantum rings \cite{Uchoa}. 


In summary, we have shown that Mott insulators having spin and orbital degeneracies can be artificially designed in a solid state system. The  emergent SU(4) symmetry of the problem follows from the unique nature of the valley degrees of freedom in honeycomb substrates and does not require fine tuning. We have predicted the conditions for Coulomb impurity lattices to be in the Mott regime and discussed experiments that could detect quantum spin-orbital liquid states. 

Most of
the current efforts to simulate quantum spin liquids are concentrated in cold atom systems, where the Mott physics is present only at ultra low temperatures  \cite{Jordens, Cazalilla}. This proposal may lead to significant advances in the experimental
design and observation of quantum spin-orbital liquids in solid-state settings.


\section*{Methods}

{\bf Wavefunctions.} 
\textcolor{black}{We assume a real space cut-off
for the Coulomb interaction $a=\lambda_{C}/18$. For a typical mass
gap energy $mv^{2}\approx0.13$ eV and $\hbar v\approx6$eV$\mbox{\AA}$,
as in graphene on SiC, the Compton wavelength $\lambda_{C}\sim50\mbox{\AA}$,
which corresponds to $a\approx2.8\mbox{\AA}$. This number agrees
with the typical size of many Coulomb impurities, including alkaline
metals.}

\textcolor{black}{The analytical form of the 2D Coulomb impurity wavefunctions
in the weak coupling regime ($g\ll1$) is well known \cite{Novikov,Kotov}.
In that regime the cutoff does not play an important role (can be
set to zero) and the bound states are shallow. The wavefunctions in
the subcritical strong coupling regime ($0.5\lesssim g<g_{c}$) can
be solved analytically as well. They correspond to the solution of
the Dirac equation in the potential (\ref{eq:Vr}) and bare strong
similarity to the 3D Dirac equation (QED$_{3+1}$) case \cite{Greiner,Zeldovich}. }

\textcolor{black}{Setting $\hbar=v=1$, for $r>a$, the strong coupling
solution in the subcritical regime has spinor component amplitudes
\cite{Supp}} 
\begin{equation}
F_{j}^{(\pm)}(r)=\sqrt{m\mp\epsilon}\, e^{-\rho/2}\rho^{-\gamma-1/2}\frac{\Gamma(2s\gamma)}{\Gamma(s\gamma-\tilde{\epsilon})}G^{(\pm)}(r),\label{eq:F}
\end{equation}
where $\gamma=\sqrt{j^{2}-g^{2}}$, $\beta=\sqrt{m^{2}-\epsilon^{2}}$,
$\Gamma(x)$ is a gamma function and 
\begin{eqnarray}
G^{(\pm)}(r) & \equiv & \sum_{s=\pm1}\left[\mathcal{F}(-\gamma-\tilde{\epsilon};1-2\gamma;\rho)\right.\nonumber \\
 &  & \qquad\left.\mp\frac{-\gamma-\tilde{\epsilon}}{j+\tilde{m}}\mathcal{F}(1-s\gamma-\tilde{\epsilon};1-2s\gamma;\rho)\right]\:\quad\label{eq:G}
\end{eqnarray}
is defined in terms of confluent hypergeometric functions of the first
kind. $\tilde{m}=mg/\beta$, $\tilde{\epsilon}=\epsilon g/\beta$
and $\rho=2\beta r$ are the normalized mass, energy and distance
away from the impurity. For $r\leq a$, the solution is defined in
terms of Bessel functions 
\begin{eqnarray}
F_{j}^{-}(r) & = & J_{j-1/2}(\sqrt{E_{+}E_{-}}r)\label{eq:f2}
\end{eqnarray}
and 
\begin{equation}
F_{j}^{(+)}(r)=-\frac{1}{E_{+}}\left\{ \partial_{r}[\sqrt{r}F_{j}^{(-)}(r)]-\frac{j}{r}\sqrt{r}F_{j}^{(-)}(r)\right\} ,\label{eq:F^+}
\end{equation}
where $E_{\pm}=\epsilon-V(a)\pm m$.

The energy of the levels follows from matching the wavefunctions at
$r=a$, $\Psi_{r<a}(a)=\Psi_{r>a}(a)$, as shown in Fig. 2. For a
given angular momentum state $j$, there is an infinite number of
solutions that can be labeled by the index $n\in\mathbb{N}$, which
is a non-negative integer. The lowest energy solution is labeled $n=0$,
with higher $n>0$ attributed to the other higher excited states.
For $j=\frac{1}{2}$ and $\epsilon=-m$, the critical coupling of
the $n=0$ level state is $g_{c}=0.916$. The spectrum is in excellent
agreement with the numerical results of \cite{Pereira}.

{\bf Hubbard \textbf{\emph{U}} term. }
The Coulomb interaction
among electrons in the lowest energy state $n=0$ and $j=\frac{1}{2}$
is described by 
\begin{equation}
\mathcal{H}_{C}=\frac{1}{2}\int\mbox{d}^{2}r\mbox{d}^{2}r^{\prime}\hat{\rho}(\mathbf{r})\frac{e^{2}}{\kappa|\mathbf{r}-\mathbf{r}^{\prime}|}\hat{\rho}(\mathbf{r}^{\prime}),\label{eq:Hc}
\end{equation}
where $\hat{\rho}(\mathbf{r})=\sum_{\sigma}\hat{\Theta}_{\sigma}^{\dagger}(\mathbf{r})\hat{\Theta}_{\sigma}(\mathbf{r})$
is the density operator defined in terms of the field operator $\hat{\Theta}_{\sigma}(\mathbf{r})=\sum_{\nu}\Phi_{\frac{1}{2},\nu}(\mathbf{r})c_{\nu,\sigma}$.
Hamiltonian (\ref{eq:Hc}) can be expressed explicitly in terms of
$c$ operators, resulting in the Hubbard $U$ Hamiltonian described
in the main text. The exchange term that also follows from (\ref{eq:Hc})
is identically zero due to the orthogonality of the two valley eigenspinors.

{\bf Spin-orbital exchange Hamiltonian. }
In second order of perturbation
theory, the superexchange Hamiltonian is expressed in terms of $c$
operators as: 
\begin{equation}
\mathcal{H}_{s}=-J_{s}\sum_{\langle ij\rangle}\sum_{\{\nu\},\{\sigma\}}c_{i,\nu,\sigma}^{\dagger}c_{j,\nu,\sigma}c_{j,\nu^{\prime},\sigma^{\prime}}^{\dagger}c_{i,\nu^{\prime},\sigma^{\prime}},\label{eq:Hs}
\end{equation}
with $J_{s}=t^{2}/U$. The exchange interaction between NN sites can
be calculated from the Coulomb interaction $\sum_{\langle ij\rangle}\mathcal{H}_{C,ij}$,
\begin{equation}
\mathcal{H}_{C,ij}=\frac{1}{2}\int\mbox{d}^{2}r\mbox{d}^{2}r^{\prime}\hat{\rho}(\mathbf{r}_{i})\frac{e^{2}}{\kappa|\mathbf{r}-\mathbf{r}^{\prime}|}\hat{\rho}(\mathbf{r}_{j}^{\prime}).\label{eq:Hc-1}
\end{equation}
We extend the definition of the field operators as a sum over lattice
sites, $\Theta_{\sigma}(\mathbf{r})=\sum_{\nu,i}\Phi_{\frac{1}{2},\nu}(\mathbf{r}_{i})c_{i,\nu\sigma}$.
The exchange part of the interaction above term can be explicitly
written as 
\begin{equation}
\mathcal{H}_{e}=J_{e}\sum_{\langle ij\rangle}\sum_{\{\nu\}\{\sigma\}}c_{i,\nu,\sigma}^{\dagger}c_{j,\nu^{\prime},\sigma^{\prime}}^{\dagger}c_{i\nu^{\prime},\sigma^{\prime}}c_{j,\nu,\sigma},\label{eq:He}
\end{equation}
where $J_{e}$ is given in the text. Hamiltonians (\ref{eq:Hs})
and (\ref{eq:He}) both map into pseudospin (valley) and spin operators,
$\boldsymbol{\tau}=(\tau^{x},\tau^{y},\tau^{z})$ and \textbf{$\mathbf{S}=(S^{x},S^{y},S^{z})$},
through the following relations: 
\begin{eqnarray*}
c_{i,\nu,\sigma}^{\dagger}c_{i,\nu,\sigma} & \to & \left(\frac{1}{2}+\nu\tau_{i}^{z}\right)\left(\frac{1}{2}+\sigma S_{i}^{z}\right)\\
c_{i,\nu,\sigma}^{\dagger}c_{i,-\nu,\sigma} & \to & \tau^{\nu}\left(\frac{1}{2}+\sigma S_{i}^{z}\right)\\
c_{i,\nu,\sigma}^{\dagger}c_{i,\nu,-\sigma} & \to & \left(\frac{1}{2}+\nu\tau_{i}^{z}\right)S^{\sigma}\\
c_{i,\nu,\sigma}^{\dagger}c_{i,-\nu,-\sigma} & \to & \tau^{\nu}S^{\sigma},
\end{eqnarray*}
where $\tau^{\nu}=\left(\tau^{x}+\nu i\tau^{y}\right)$ and $S^{\sigma}=S^{x}+\sigma iS^{y}$.
$\nu=\pm$, and $\sigma=\pm$ indexes the two valleys and spins respectively.
This mapping results in Hamiltonian (\ref{eq:JHam}).

\section*{Acknowledgements}

We are grateful to Kevin Beach, Frederic Mila and Kieran Mullen for numerous
stimulating discussions. X. Dou and B. Uchoa acknowledge the University
of Oklahoma for partial support. B. Uchoa was supported by NSF CAREER
grant NMR-1352604. V. N. Kotov acknowledges support by the U.S. Department
of Energy (DOE) grant DE-FG02-08ER46512. B. Uchoa thanks the Aspen
Center of Physics where this work was partially completed.

\section*{Author contributions}

X. D.  did the analytical and numerical calculations. V. N. K. performed calculations and commented on the manuscript.  B. U. coordinated the project, performed analytical calculations and wrote the manuscript. 

\section*{Additional information}

{\bf Competing financial interests:} The authors declare no competing financial interests.

\section*{Correspondence}

Correspondence and requests for materials should be addressed to B. U. (uchoa@ou.edu).

\end{document}


\title{Suplementary Materials of ``Designing Quantum Spin-Orbital
Liquids in Artificial Mott Insulators''}

\author{Xu Dou}

\affiliation{Department of Physics and Astronomy, University of Oklahoma, Norman,
OK 73069, USA}

\author{Valeri N. Kotov}

\affiliation{Department of Physics, University of Vermont, Burlington, VT 05405,
USA}

\author{Bruno Uchoa}

\email{uchoa@ou.edu}

\selectlanguage{english}%

\affiliation{Department of Physics and Astronomy, University of Oklahoma, Norman,
OK 73069, USA}

\maketitle

\date{\today}

\section*{Wavefunction of the strong coupling subcritical regime}

The wave function $\Psi(\mathbf{r})$ of a two dimensional massive
Dirac fermion moving around a Coulomb impurity satisfies,
\begin{equation}
(-i\boldsymbol{\sigma}\cdot\boldsymbol{\nabla}+V(r)+m\sigma_{z})\Psi(\mathbf{r})=\epsilon\Psi(\mathbf{r}),
\end{equation}
where
\begin{align*}
V(r) & =\begin{cases}
-g/r, & r>a\\
-g/a, & r\leq a
\end{cases}
\end{align*}
is the regularized Coulomb potential. Here we set $\hbar=v=1$. 

The two-component wave function is
\begin{equation}
\Psi(r,\phi)=\frac{1}{\sqrt{2\pi}}\left(\begin{array}{c}
F_{j}^{(-)}(r)e^{i(j-1/2)\phi}\\
F_{j}^{(+)}(r)e^{i(j+1/2)\phi}
\end{array}\right),\label{eq:2}
\end{equation}
he Dirac equation with the presence of an external potential becomes
\begin{align*}
\left[\begin{array}{cc}
\epsilon-m-U & -(\partial_{r}+\frac{\kappa+1}{r})\\
(\partial_{r}-\frac{\kappa}{r}) & \epsilon+m-U
\end{array}\right]\left(\begin{array}{c}
F_{j}^{(-)}(r)\\
F_{j}^{(+)}(r)
\end{array}\right) & =0
\end{align*}
or equivalently
\begin{align}
\frac{dF_{j}^{(-)}}{dr}-\frac{\kappa}{r}F_{j}^{(-)}+(\epsilon+m-U)F_{j}^{(+)} & =0\nonumber \\
\frac{dF_{j}^{(+)}}{dr}+\frac{\kappa+1}{r}F_{j}^{(+)}-(\epsilon-m-U)F_{j}^{(-)} & =0\label{eq:3}
\end{align}
where $\kappa=j-\frac{1}{2}$.

\subsection{Solution for $r>a$}

The wave functions assumes the form ($j$ indexes will be dropped)
\[
F^{(-)}(\rho)=\sqrt{m+\epsilon}e^{-\rho/2}\rho^{\gamma-1/2}(Q_{1}+Q_{2})
\]
\begin{equation}
F^{(+)}(\rho)=\sqrt{m-\epsilon}e^{-\rho/2}\rho^{\gamma-1/2}(Q_{1}-Q_{2}).\label{eq:f}
\end{equation}
where $\rho=2\lambda r$, $\beta=\sqrt{m^{2}-\epsilon^{2}}$, and$\gamma=\sqrt{j^{2}-g^{2}}$.
After some algebra, we get

\begin{equation}
\text{\ensuremath{\rho}}Q_{1}^{\prime}+(\gamma-\frac{\epsilon g}{\beta})Q_{1}-(j+\frac{mg}{\beta})Q_{2}=0\label{eq:p12}
\end{equation}
\begin{equation}
\rho Q_{2}^{\prime}+(\gamma+\frac{\epsilon g}{\beta}-\rho)Q_{2}-(j-\frac{mg}{\beta})Q_{1}=0.
\end{equation}

These equations can be decoupled 
\[
\rho Q_{1}^{\prime\prime}+(1+2\gamma-\rho)Q_{1}^{\prime}-(\gamma-\frac{\epsilon g}{\beta})Q_{1}=0
\]
\[
\rho Q_{2}^{\prime\prime}+(1+2\gamma-\rho)Q_{2}^{\prime}-(1+\gamma-\frac{\epsilon g}{\beta})Q_{2}=0,
\]
both of which are Kummer's differential equation
\[
xy^{\prime\prime}+(c-x)y^{\prime}-ay=0
\]
with the solution $y=c\mathcal{F}(a;c;x)$. Here we call the confluent
Hypergeometric function of the first kind $_{1}F_{1}(a;c;x)$ as $\mathcal{F}(a;c;x)$,

\[
_{1}F_{1}(a;c;x)=1+\frac{a}{c}x+\frac{a(a+1)}{c(c+1)}\frac{x^{2}}{2!}+...,
\]
and notice that $_{1}F_{1}(a;c;x=0)=1.$ The solutions are 
\begin{equation}
Q_{1}=c_{1}\mathcal{F}(\gamma-\frac{\epsilon g}{\beta};1+2\gamma;\rho)+d_{1}\mathcal{F}(-\gamma-\frac{\epsilon g}{\beta};1-2\gamma;\rho)\label{eq:q1}
\end{equation}
\begin{equation}
Q_{2}=c_{2}\mathcal{F}(1+\gamma-\frac{\epsilon g}{\beta};1+2\gamma;\rho)+d_{2}\mathcal{F}(1-\gamma-\frac{\epsilon g}{\beta};1-2\gamma;\rho).\label{eq:q2}
\end{equation}

\subsubsection{Weak coupling regime}

When $g<j$, $\gamma$ is real. Integrability of the wavefunction
at $\rho\to\infty$ requires that $d_{1}=d_{2}=0$. From Eq.$(\ref{eq:p12})$,
one can determine the ratio 
\[
\frac{c_{1}}{c_{2}}=\frac{Q_{1}}{Q_{2}}\bigg|{}_{\rho=0}=\frac{j+\frac{mg}{\beta}}{\gamma-\frac{\epsilon g}{\beta}}.
\]
To simplify the notation we call $\tilde{m}=mg/\beta$, $\tilde{\epsilon}=\epsilon g/\beta$,
and 
\[
c_{1}=c,\ c_{2}=\frac{\gamma-\tilde{\epsilon}}{j+\tilde{m}}c.
\]
That leads to the solution 
\[
F_{j}^{(-)}(\rho)=c\sqrt{m+\epsilon}e^{-\rho/2}\rho^{\gamma-1/2}[\mathcal{F}(\gamma-\tilde{\epsilon};1+2\gamma;\rho)+\frac{\gamma-\tilde{\epsilon}}{j+\tilde{m}}\mathcal{F}(1+\gamma-\tilde{\epsilon};1+2\gamma;\rho)]
\]
\[
F_{j}^{(+)}(\rho)=c\sqrt{m-\epsilon}e^{-\rho/2}\rho^{\gamma-1/2}[\mathcal{F}(\gamma-\tilde{\epsilon};1+2\gamma;\rho)-\frac{\gamma-\tilde{\epsilon}}{j+\tilde{m}}\mathcal{F}(1+\gamma-\tilde{\epsilon};1+2\gamma;\rho)]
\]
This solution is regular at $\rho\to0$, and the short distance cut-off
can be set to zero.

\subsubsection{Strong coupling regime}

When the coupling $g>\frac{1}{2}$, $\gamma$ becomes imaginary. With
the requirement of imposing a small distance cut-off, the condition
that the wave function behaves well at $\rho=0$ is not necessary,
so we should include both $\pm\gamma$ branches into the solutions.
The ratio between the two And in this case we hope the wave functions
die off at $\rho\rightarrow+\infty$, which can also serve to settle
down the ratio between $\gamma-$branch and $(-\gamma)$-branch. The
formula can be used here is the asymptotic form of the hypergeometric
function, 
\[
_{1}F_{1}(a;b;x)\sim\Gamma(b)\left(\frac{e^{z}z^{a-b}}{\Gamma(a)}+\frac{(-z)^{-a}}{\Gamma(b-a)}\right).
\]
The second part is required if the gamma function $\Gamma(a)$ is
infinite (when $a$ is a negative integer) or $\mathrm{Re}(z)$ is
non-positive. In our case, we could exclude these two conditions,
and only keep the second term. Therefore for large $|z|$, the dominating
part (which is growing) of $\mathcal{F}(z)$ is 
\[
\mathcal{F}(a;b;z)\sim\frac{\Gamma(b)}{\Gamma(a)}e^{z}z^{a-b},
\]
and we ask for some condition to cancel this term. For $aF(\rho;\gamma)+bF(\rho;-\gamma)$
we need the following two terms to be finite at $\rho\rightarrow+\infty$

\begin{equation}
a\rho^{\gamma-1/2}\mathcal{F}(\gamma-\tilde{\epsilon};1+2\gamma;\rho)+b\rho^{-\gamma-1/2}\mathcal{F}(-\gamma-\tilde{\epsilon};1-2\gamma;\rho)\label{eq:c1}
\end{equation}
\begin{equation}
a\rho^{\gamma-1/2}\frac{\gamma-\tilde{\epsilon}}{j+\tilde{m}}\mathcal{F}(1+\gamma-\tilde{\epsilon};1+2\gamma;\rho)+b\rho^{-\gamma-1/2}\frac{-\gamma-\tilde{\epsilon}}{j+\tilde{m}}\mathcal{F}(1-\gamma-\tilde{\epsilon};1-2\gamma;\rho)\label{eq:c2}
\end{equation}
From $\ref{eq:c1}$, 
\[
\frac{a}{b}=-\frac{\Gamma(\gamma-\tilde{\epsilon})}{\Gamma(1+2\gamma)}\frac{\Gamma(1-2\gamma)}{\Gamma(-\gamma-\tilde{\epsilon})}=-\frac{\Gamma(\gamma-\tilde{\epsilon})}{(2\gamma)\Gamma(2\gamma)}\frac{(-2\gamma)\Gamma(-2\gamma)}{\Gamma(-\gamma-\tilde{\epsilon})}=\frac{\Gamma(\gamma-\tilde{\epsilon})}{\Gamma(2\gamma)}\frac{\Gamma(-2\gamma)}{\Gamma(-\gamma-\tilde{\epsilon})}.
\]
We can assign 
\begin{equation}
a=\frac{\Gamma(-2\gamma)}{\Gamma(-\gamma-\tilde{\epsilon})},\qquad\qquad b=\frac{\Gamma(2\gamma)}{\Gamma(\gamma-\tilde{\epsilon})}\label{eq:ab}
\end{equation}

The solution for $F^{(\pm)}$ is 
\begin{align}
F_{j}^{(\mp)}(r) & =c^{\prime}\sqrt{m\pm\epsilon}e^{-\rho/2}\rho^{-1/2}\nonumber \\
 & \times\left[\frac{\Gamma(-2\gamma)}{\Gamma(-\gamma-\tilde{\epsilon})}\rho^{\gamma}\mathcal{F}(\gamma-\tilde{\epsilon};1+2\gamma;\rho)+\frac{\Gamma(2\gamma)}{\Gamma(\gamma-\tilde{\epsilon})}\rho^{-\gamma}\mathcal{F}(-\gamma-\tilde{\epsilon};1-2\gamma;\rho)\right.\nonumber \\
 & \left|\pm\frac{\Gamma(-2\gamma)}{\Gamma(-\gamma-\tilde{\epsilon})}\frac{\gamma-\tilde{\epsilon}}{j+\tilde{m}}\rho^{\gamma}\mathcal{F}(1+\gamma-\tilde{\epsilon};1+2\gamma;\rho)\pm\frac{\Gamma(2\gamma)}{\Gamma(\gamma-\tilde{\epsilon})}\frac{-\gamma-\tilde{\epsilon}}{j+\tilde{m}}\rho^{-\gamma}\mathcal{F}(1-\gamma-\tilde{\epsilon};1-2\gamma;\rho)\right]\label{eq:solution}
\end{align}

\subsection{Solution for $r\leq a$}

In the $r<a$ region, we define 
\begin{align*}
F_{j}^{(-)}(r) & =\frac{1}{\sqrt{r}}A(r)\\
F_{j}^{(+)}(r) & =\frac{1}{\sqrt{r}}B(r)
\end{align*}
the Dirac equation becomes
\[
A^{\prime}(r)-\frac{j}{r}A(r)+E_{+}B(r)=0
\]
\[
B^{\prime}(r)+\frac{j}{r}B(r)-E_{-}A(r)=0
\]
where $E_{\pm}=\epsilon+\frac{g}{a}\pm m$. These equations can be
decoupled into 
\[
A^{\prime\prime}(r)+(E_{+}E_{-}+\frac{j-j^{2}}{r^{2}})A(r)=0
\]
\[
B^{\prime\prime}(r)+(E_{+}E_{-}-\frac{j+j^{2}}{r^{2}})B(r)=0
\]

The solutions are 
\[
A(r)=c_{1}\sqrt{r}J_{j-1/2}(\sqrt{E_{+}E_{-}}r)
\]

\[
B(r)=c_{2}\sqrt{r}J_{j+1/2}(\sqrt{E_{+}E_{-}}r)
\]
and $\sqrt{E_{+}E_{-}}=\sqrt{\epsilon^{2}+(g/a)^{2}+(2\epsilon g/a)-m^{2}}$.

From 
\[
B(r)=-\frac{A^{\prime}-\frac{j}{r}A}{E_{+}},
\]
we have 
\[
B(r)=-\frac{c_{1}}{E_{+}}[\frac{\frac{1}{2}-j}{\sqrt{r}}J_{j-1/2}(\sqrt{E_{+}E_{-}}r)+\sqrt{r}J_{j-1/2}^{\prime}(\sqrt{E_{+}E_{-}}r)]
\]

\subsection{Energy}

The energy $\epsilon$ can be determined by matching the inside solution
and the outside one, formally through 
\[
Out(j,\epsilon,r,g)\big|_{r=a}=Ins(j,\epsilon,r,g)\big|_{r=a}
\]
For given $j$, $g$, and at $r=a$, we can determine the energy $\epsilon$

\[
\left(\begin{array}{c}
\sqrt{r}F(r)\\
\sqrt{r}G(r)
\end{array}\right)\bigg|_{r=a}=\left(\begin{array}{c}
A(r)\\
B(r)
\end{array}\right)\bigg|_{r=a}
\]